\documentclass[twocolumn,showpacs,preprintnumbers,amsmath,amssymb]{revtex4}
\usepackage{graphicx}
\usepackage{dcolumn}
\usepackage{latexsym}
\usepackage{dsfont}
\usepackage{bm}
\begin{document} 
\title{Energy density for chiral lattice fermions with chemical potential}

\author{Christof Gattringer$^a$} 
\author{Ludovit Liptak$^b$}

\affiliation{\vspace{4mm}$^a$Institut f\"ur Physik, FB Theoretische Physik,
Universit\"at Graz,
\vskip0mm
Universit\"atsplatz 5, 8010 Graz, Austria
\vskip1mm       
$^b$Institute of Physics, Slovak Academy of Sciences,
\vskip0mm
D\'ubravsk\'a cesta 9, 845 11 Bratislava 45, Slovak Republic}


\begin{abstract}
We study a recently proposed formulation of overlap fermions
at finite density. In particular we compute the energy density
as a function of the chemical potential and the temperature.  
It is shown that overlap fermions with chemical potential
approach the correct continuum behavior. 
\end{abstract}

\maketitle

\section{Introduction}

Over the last two decades lattice gauge theory was turned
into a powerful qualitative tool for analyzing QCD. This 
progress is in part due to the advances in algorithms and 
computer technology, but also on the conceptual side 
important breakthroughs were made. Most prominent among these is 
the correct implementation of chiral symmetry on 
the lattice based on the Ginsparg-Wilson equation for
the Dirac operator \cite{giwi}. 

An application of lattice techniques which has seen a lot of attention 
in recent years, is the study of QCD at finite temperature.  
The lattice implementation of the chemical potential $\mu$, 
necessary for such an
analysis, is not straightforward, however. It is well known \cite{haka}, 
that a naive introduction leads to $\mu^2/a^2$ contributions which 
diverge in the continuum limit when the lattice spacing $a$ is sent 
to zero. For more traditional formulations, such as the Wilson 
or staggered Dirac operators, the problem has been solved by
introducing the chemical potential in the same way as
the 4-component of the gauge field.

A satisfactory implementation of the chemical potential 
should be compatible with chiral symmetry on the lattice based 
on the Ginsparg-Wilson equation. When attempting to introduce 
the chemical potential into the only solution of the 
Ginsparg-Wilson equation know in closed form, 
the overlap operator \cite{overlap},
a potential problem quickly surfaces: defining the sign function
of a non-hermitian matrix. In \cite{blwe} Bloch and Wettig
proposed a solution based on an analytic continuation of the sign 
function into  the complex plane. It was shown, that the eigenvalue 
spectra of this construction match the 
expectations from random matrix theory. 

In this letter we analyze the proposal \cite{blwe} further and 
study the energy density of free, massless overlap fermions with 
chemical potential. The dependence of the energy density 
on $\mu$ and the temperature $T$ allows for a detailed analysis 
of the lattice formulation at finite density.  
Of particular interest will be the 
question whether the analytic continuation of the sign function 
produces divergent $\mu^2/a^2$ terms. Our study indicates the absence of such
contributions and we find that the $\mu$ and $T$ dependence of the 
energy density
is approached correctly. 

\section{Setup of the calculation}

The overlap Dirac operator $D(\mu)$ for
fermions with a chemical potential $\mu$ is given as
\begin{eqnarray}
D(\mu) & = & \frac{1}{a} [ 1 - \gamma_5 \, \mbox{sign}\, H(\mu) ] \; ,
\nonumber
\\
H(\mu) & = & \gamma_5 \, [ 1 - a D_W(\mu)] \; . 
\label{overlap}
\end{eqnarray}
The sign function may be defined through the spectral theorem for matrices.
$D_W(\mu)$ denotes the usual Wilson Dirac operator,
\begin{eqnarray}
& & 
\!\!\!\!\!\!
D_W\!(\mu)_{x,y} \, = \, \mathds{1} \Big[\frac{3}{a} + \frac{1}{a_4} \Big] \, 
 \delta_{x,y} \; -
\\  
& & 
\!\!\!\!\!\!
\sum_{j = 1}^{3} \Big[ 
\frac{\mathds{1}\! - \!\gamma_j}{2a} 
U_j(x) \delta_{x+\hat{j},y} \, + \,  
\frac{\mathds{1} \!+ \!\gamma_j}{2a}
U_j(x\!-\!\hat{j})^\dagger \delta_{x-\hat{j},y} \Big] \, -  
\nonumber \\
& & 
\!\!\!\!\!\!
\frac{\mathds{1} \! - \!\gamma_4}{2a_4} 
U_4(x) e^{\mu a_4} \delta_{x+\hat{4},y} \, - \, 
\frac{\mathds{1} \!+\! \gamma_4}{2a_4}
U_4(x\!-\!\hat{4})^\dagger  e^{-\mu a_4} \delta_{x-\hat{4},y} \; . 
\nonumber  
\end{eqnarray}
For later use we distinguish between the lattice spacing $a$ 
in spatial direction and the temporal lattice constant $a_4$.
Periodic boundary conditions are used in the spatial directions,
while in time direction we apply anti-periodic boundary conditions.  
The chemical potential $\mu$ is coupled in the usual exponential form
\cite{haka}.

For vanishing $\mu$ the Wilson Dirac operator is $\gamma_5$-hermitian,
i.e.,\ $\gamma_5 D_W(0) \gamma_5 = D_W(0)^\dagger$. This implies that $H(0)$
is a hermitian matrix. As soon as the chemical 
potential $\mu$ is turned on, $\gamma_5$-hermiticity no longer holds, and 
$H(\mu)$ is a non-hermitian, general matrix. This fact has two important 
consequences: Firstly, the eigenvalues of $H(\mu)$ are no longer real and the 
sign function for a complex number has to be defined in the spectral 
representation of sign$\,H(\mu)$. Secondly, the spectral representation has
to be formulated using left and right eigenvectors. This latter problem 
will be dealt with later when we discuss the evaluation of sign$\,H(\mu)$.
For the sign function of a complex number we use the analytic continuation 
proposed in \cite{blwe} and define the sign function through the sign of 
the real part
\begin{equation}
\mbox{sign}\, (x + i y) \; = \; \mbox{sign} \, (x) \; .
\end{equation}

The observable we study here is the energy density defined as
\begin{eqnarray}
&& \epsilon(\mu) \; = \; \frac{1}{V} \langle {\cal H} \rangle \; = \;
\frac{1}{V} \frac{ \mbox{Tr} \Big[ {\cal H} \, e^{-\beta\, 
( {\cal H} - \mu {\cal N})}\Big] }{Z} \; =
\label{epsidef}
\\
&& - \frac{1}{V} \frac{\partial}{\partial \beta} 
\ln \mbox{Tr} \Big[ e^{-\beta ( {\cal H} - \mu {\cal N})}\Big]_{\beta\mu = c}
\; = \; - \frac{1}{V} \frac{\partial \ln Z}{\partial \beta}
\bigg|_{\beta\mu = c} \; .
\nonumber
\end{eqnarray}
Here ${\cal H}$ is the Hamiltonian of the system, ${\cal N}$ 
denotes the number 
operator and $\beta = 1/T$ is the inverse temperature (in our units the 
Boltzmann constant $k$ is set to $k=1$). The derivatives 
in the second line are taken such that $\beta\mu = 
c =$ const. 

The continuum result for the subtracted energy density of
free massless fermions reads
(see, e.g., \cite{kapusta})
\begin{equation}
\epsilon(\mu) - \epsilon(0) \; = \; \frac{\mu^4}{4\pi^2} \; + \; 
\frac{1}{2} \mu^2 T^2 \; .
\label{contresult}
\end{equation}

When working on the lattice, the inverse temperature $\beta$ is given by
the lattice extent in 4-direction, i.e., $\beta = N_4 a_4$. Thus the
derivative $\partial/\partial \beta$ in (\ref{epsidef}) turns into
$N_4^{-1} \partial/\partial a_4$. The partition function $Z$ is given 
by the fermion determinant $\det D$ which we write as
the product over all eigenvalues $\lambda_n$. We thus find
\begin{eqnarray}
\epsilon(\mu) &\! = \! & - \frac{1}{V N_4}
\frac{\partial \ln \det D}{\partial a_4}\bigg|_{a_4\mu = c}\!\!\!\! \!\!= \; 
- \frac{1}{V N_4}
\frac{\partial \ln \prod_n \lambda_n}{\partial a_4}\bigg|_{a_4\mu = c}
\nonumber \\
&\! = \! & - \frac{1}{V N_4} \sum_n \frac{1}{\lambda_n} 
\frac{\partial \lambda_n}{\partial a_4}\bigg|_{a_4\mu = c}\; .
\label{specsum}
\end{eqnarray} 

\section{Evaluation of the eigenvalues}

According to (\ref{specsum}), for the evaluation of $\epsilon(\mu)$ 
the eigenvalues $\lambda_n$ of the Dirac operator $D$ have to be 
computed. This is done in three steps: First we bring the Dirac 
operator for free fermions to $4\times4$ block-diagonal form, using Fourier
transformation. Subsequently the spectral representation is applied 
to the $4\times4$  blocks of $H$ to evaluate sign $H$. Finally 
the eigenvalues of the blocks of $D$ are computed and by summing 
over the discrete momenta all eigenvalues are obtained.

Following this strategy, one finds for the Fourier transform 
$\widehat{H}$ of $H$,
\begin{equation}
\widehat{H} \; = \; \gamma_5 h_5 \; + \; i \gamma_5 \sum_\nu \gamma_\nu h_\nu 
\; ,
\end{equation}
with
\begin{eqnarray}
h_5 & = & 1 - \sum_{j=1}^3 [ 1 - \cos(a p_j)] - 
\frac{a}{a_4} [ 1 - \cos(a_4 (p_4 - i\mu))] \; ,
\nonumber \\
h_j & = & - \sin(a p_j) \quad \mbox{for} \quad j = 1,2,3 \; ,
\nonumber \\
h_4 & = & - \frac{a}{a_4} \sin(a_4 (p_4 - i\mu)) \; .
\label{hdefs}
\end{eqnarray}

The spatial momenta are given by $p_j =  2\pi k_j/aN$, where $N$ is the number 
of lattice points in the spatial directions and $k_j = 0,1\, ...\, N-1$.  
The momenta in time-direction are $p_4 =  \pi( 2k_4 + 1)/a_4 N_4$, 
$k_4 = 0,1\, ...\, N_4-1$.

The remaining diagonalization of $\widehat{H}$ is similar to the construction 
of the left- and right-eigenfunctions for the free Dirac operator. One finds that 
$\widehat{H}$ has two different, doubly degenerate eigenvalues 
\begin{equation}
\alpha_1 = \alpha_2 \, = \, + \, s \; , \; 
\alpha_3 = \alpha_4 \, = \, - \, s \; , \; s \, = \, \sqrt{h^2 + h_5^2} \; ,
\end{equation}
where $h^2 = \sum_\nu h_\nu^2$. The corresponding left- and right-eigenvectors,
$l_j$ and $r_j$ are given by
\begin{eqnarray}
l_1 & = & l_1^{(0)} [ \widehat{H} + s \mathds{1} ] \; , \;   
l_2 = l_2^{(0)} [ \widehat{H} + s \mathds{1} ] \; , 
\nonumber \\
l_3 & = & l_3^{(0)} [ \widehat{H} - s \mathds{1} ] \; , \;   
l_4 = l_4^{(0)} [ \widehat{H} - s \mathds{1} ] \; , 
\nonumber \\
r_1 & = & [ \widehat{H} + s \mathds{1} ] r_1^{(0)}\; , \;   
r_2  =  [ \widehat{H} + s \mathds{1} ] r_2^{(0)}\; , \;   
\nonumber \\
r_3 & = & [ \widehat{H} - s \mathds{1} ] r_3^{(0)}\; , \;   
r_4  =  [ \widehat{H} - s \mathds{1} ] r_4^{(0)}\; .   
\end{eqnarray}
The constant spinors $l_j^{(0)}, r_j^{(0)}$ are 
($T$ is transposition)
\begin{eqnarray}
l_1^{(0)} &\! = & r_1^{(0)\, T} \, = \, c \, (1,0,0,0) \; , \;
l_2^{(0)} \, = \, r_2^{(0)\, T} \, = \, c \, (0,1,0,0) \, , 
\nonumber \\
l_3^{(0)} &\! = & r_3^{(0)\, T} \, = \, c \,(0,0,1,0) \; , \;
l_4^{(0)} \, = \, r_4^{(0)\, T} \, = \, c \, (0,0,0,1) \, .
\nonumber 
\\
\end{eqnarray} 
The constant $c = (2 s(s+h_5))^{-1/2}$ ensures the correct normalization, 
such that the eigenvectors obey $l_i r_j = \delta_{ij}$.
 
Using these eigenvectors and the spectral theorem we find for 
sign $\widehat{H}$ the simple result
\begin{equation}
\mbox{sign} \, \widehat{H} \; = \; 
\sum_{j=1}^4 \mbox{sign}\,(\lambda_j) \, r_j \, l_j \; = \;
\frac{\mbox{sign}(s)}{s} \, \widehat{H} \; .
\end{equation}
Plugging this back into the overlap formula (\ref{overlap}) 
and diagonalizing the remaining $4\times 4$ problem one finds two 
different eigenvalues for the overlap operator at a given momentum, 
\begin{equation}
\lambda_\pm \; = \; \frac{1}{a}\left[ 1 - 
\frac{ \mbox{sign}\,(\sqrt{h^2 + h_5^2}\,)\, h_5 \pm i 
\sqrt{h^2}}{\sqrt{h^2 + h_5^2}} \right] \; ,
\label{evals}
\end{equation}
where each of the two 
eigenvalues is twofold degenerate. The momentum dependence
enters through the components $h_\nu, h_5$ defined in (\ref{hdefs}). 
In the spectral sum (\ref{specsum}) the label $n$ runs over all
momenta and the eigenvalues at fixed momentum as given in (\ref{evals}).
The necessary derivative with respect to $a_4$ is straightforward to
compute in closed form, and the spectral sum (\ref{specsum}) can then
be summed numerically. The argument of the sign function cannot become
purely imaginary
on a finite lattice, and no $\delta$-like terms occur. 
We remark, that after taking the derivative with respect to $a_4$, 
we set $a = a_4 = 1$, i.e., all the
results we present are in lattice units. 
 
\section{Results}

We begin the discussion of our results with Fig.\ 1, where we show
the subtracted energy density $\epsilon(\mu) - \epsilon(0)$ 
as a function of $\mu^4$ for three different lattice volumes. For those
lattices all 4 sides have equal length, i.e., in the thermodynamic limit
they correspond to zero temperature. Thus, according to 
(\ref{contresult}), we expect the data (symbols in Fig.\ 1)
to approach the continuum form $\mu^4/4\pi^2$ (dashed line) 
as the 4-d volume is sent to infinity. 

\begin{figure}[t]
\begin{center}
\includegraphics[width=85mm,clip]{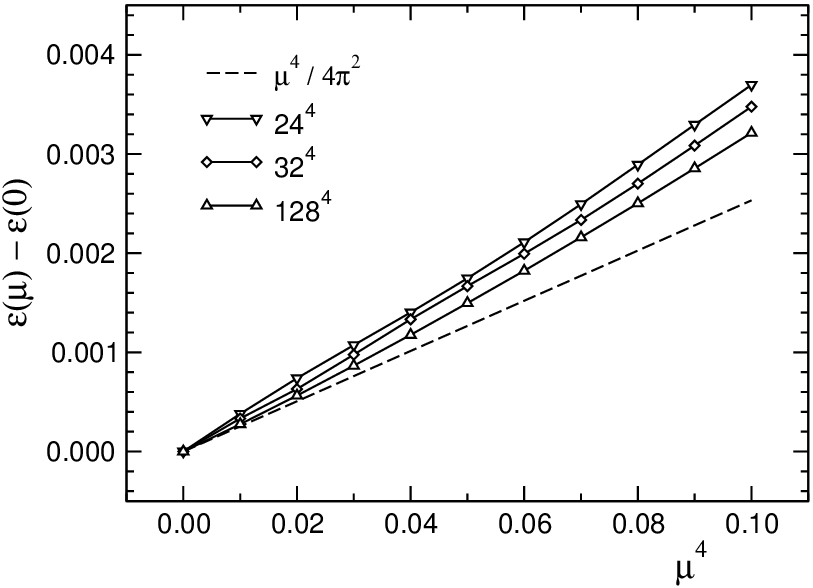} 
\end{center}
\caption{
The energy density $\epsilon(\mu) - \epsilon(0)$ as a function of
$\mu^4$ (all in lattice units). 
The symbols (connected to guide the eye) are for various
lattice sizes, the dashed line is the continuum result.}
\end{figure}

The figure clearly shows that the lattice data are predominantly linear 
when plotted versus $\mu^4$ and that for small $\mu$
they approach the continuum curve
when the volume is increased. It is, however, obvious that also on our
largest lattice still a discrepancy remains for larger $\mu$. 
In particular one finds
a slight curvature upwards, a discretization effect which here, since
the lattice spacing is just the inverse lattice extension, is also
a finite size effect. Furthermore, for small $\mu$ one expects to see finite 
temperature corrections according to (\ref{contresult}). 

\begin{figure}[t]
\begin{center}
\includegraphics[width=85mm,clip]{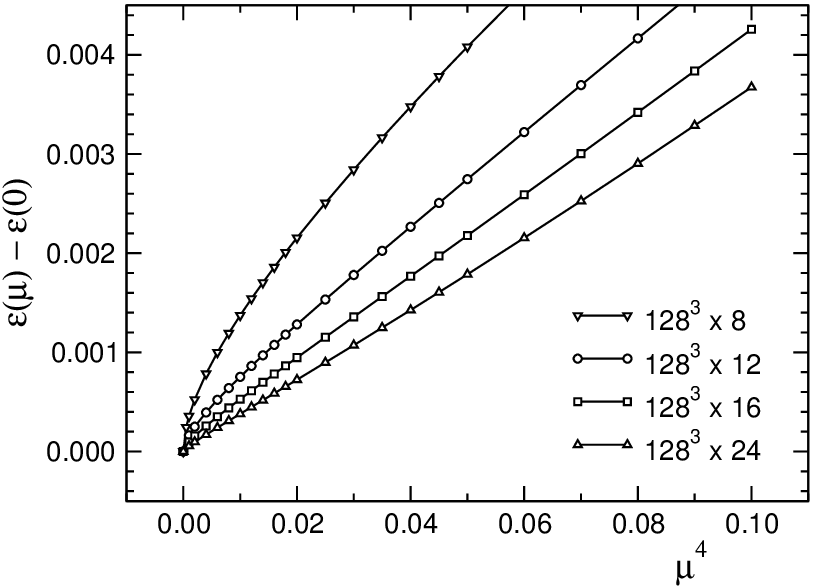} 
\end{center}
\caption{
The energy density $\epsilon(\mu) - \epsilon(0)$ as a function of
$\mu^4$, now for finite temperature lattices (all in lattice units).}
\end{figure}

In order to study these finite temperature corrections systematically, 
we analyzed lattices with short temporal extent, i.e., lattices 
with non-vanishing temperature. Fig.~2 shows the corresponding results,
where we again plot the subtracted energy density as a function of
$\mu^4$.

The lattice with the shortest temporal extent, $128^3 \times 8$, 
which corresponds to the largest temperature, shows a clear curvature.
This curvature is due to the $T^2 \mu^2 /2$ term in (\ref{contresult}),
which appears as a square root when plotted as function of $\mu^4$. 
The effect is visible also for the other lattices, but becomes
less pronounced as the temporal extent is increased, i.e., the temperature
$T$ is lowered. In order to study this effect quantitatively, we fit the
finite temperature results to the continuum form (\ref{contresult}) 
plus two terms even in $\mu$ which parameterize the cutoff effects
observed in Fig.\ 1. The fit function is given by
\begin{equation}
c_2 \, \mu^2 \, + \, c_4 \, \mu^4 \, + \, 
c_6 \, \mu^6 \, + \, c_8 \, \mu^8 \; .
\label{fitform}
\end{equation}
Due to (\ref{contresult})
the coefficient of the quadratic term should scale with the 
temperature such that one expects
\begin{equation}
c_2 \; \sim \; T^2/2 \; = \; N_4^{-2}/2 \; .
\end{equation}
The coefficient for the quartic term should be constant,
\begin{equation}
c_4 \; \sim \; 1/4\pi^2 \; = \; 0.02533 \; .
\end{equation}
The results of the fit for the data used in Fig.\ 2, and for the largest 
lattice of Fig.\ 1 are given in Table 1.

\renewcommand{\arraystretch}{1.3}
\begin{table}[b]
\begin{center}
\begin{tabular}{c c c c c c}
\hline
\hline
\;\;$N_4$\;\; &\;\; $N_4^{-2}/2$ \;\; 
& \;\;\;\;\; $c_2$\;\;\;\;\;\; 
& \;\;\;\;\; $c_4$\;\;\;\;\; 
& \;\;\;\;\; $c_6$\;\;\;\;\; 
& \;\;\;\;\; $c_8$\;\;\;\;\; \\
\hline
8   & 0.007812 & 0.010125 & 0.03519 & 0.010 &  -0.021 \\
12  & 0.003472 & 0.004125 & 0.03178 & 0.023 &  -0.013 \\
16  & 0.001953 & 0.002192 & 0.02803 & 0.029 &  -0.015 \\
24  & 0.000868 & 0.000947 & 0.02587 & 0.025 &  -0.030 \\
128 & 0.000030 & 0.000032 & 0.02543 & 0.015 & \;0.016 \\
\hline
\hline
\end{tabular}
\end{center}
\caption{
Results of the fits to the form (\ref{fitform}). The spatial volume
is always $128^3$. The temporal extension $N_4$ 
is given in the first column.
In the second column we list the corresponding value of $N_4^{-2}/2$
which is what one expects for the fitting coefficient $c_2$ in 
the third column. The coefficient $c_4$ in the fourth column is 
expected to approach the constant value $1/4\pi^2 \; = \; 0.02533$.}
\end{table}

\begin{figure}[t]
\begin{center}
\includegraphics[width=85mm,clip]{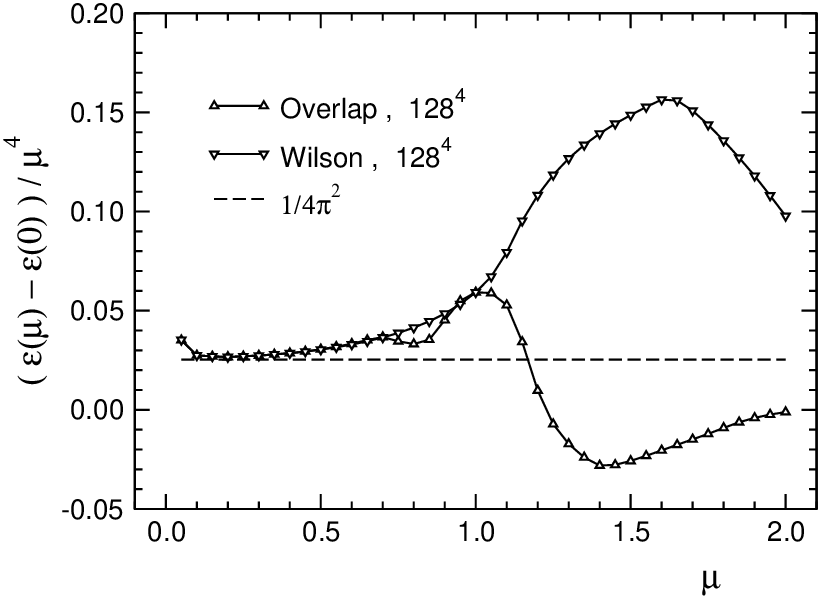} 
\end{center}
\caption{
The ratio $(\epsilon(\mu) - \epsilon(0))/\mu^4$ as a function of
$\mu$ (in lattice units). We compare the results for 
overlap to those from 
Wilson fermions.}
\end{figure}

The table shows that with increasing $N_4$ the two physically significant 
parameters $c_2$ and $c_4$ approach the values expected from the
continuum formula (\ref{contresult}): $c_2$  gets closer to $N_4^{-2}/2$
as listed in the second column, and $c_4$ approaches
$1/4\pi^2 \; = \; 0.02533$. For the largest finite temperature lattice 
$128^3 \times 24$ the discrepancy is down to 9 \% for $c_2$, and 2 \%
for $c_4$. The larger discrepancy for small $N_4$ can be understood 
as a discretization effect, since the temporal lattice spacing
$a_4$ is related to the temporal extension through $a_4 = 1/N_4$ 
and thus larger $N_4$ implies a smaller $a_4$. 
For comparison we also display the fit results for
the $128^4$ lattice, which corresponds to zero temperature.
There we find excellent agreement (less than 1\% discrepancy)
for the parameter $c_4$, governing the leading term 
at $T = 0$. The overall picture
obtained from the fit results is that overlap fermions with chemical 
potential reproduce very well both, the $\mu^4$ term, as well as the finite
temperature contribution $T^2 \mu^2 /2$. We conclude that
the analytic continuation of the sign function does
not introduce lattice artifacts, such as the $\mu^2/a^2$ term 
known to be present in a naive implementation of the chemical potential. 

In the final step of our analysis we study 
the discretization effect for larger values 
of $\mu$ and compare the results to the data from the standard Wilson
operator. In Fig.\ 3 we plot the ratio $(\epsilon(\mu) - \epsilon(0))/\mu^4$
as a function of $\mu$. In the continuum at $T = 0$ this 
ratio has the value $1/4\pi^2 \; = \; 0.02533$ indicated by 
the horizontal line. For small $\mu$, up to about 
$\mu \sim 0.7$, the Wilson and overlap data fall on top 
of each other. For very small $\mu$ both operators show a prominent 
increase which is a left-over finite temperature effect, which 
for the ratio $(\epsilon(\mu) - \epsilon(0))/\mu^4$ shows up as
a $1/\mu^2$ term. In the range between $\mu = 0.1$ and 0.5 
the data are close to the continuum value. Beyond 0.5 the 
discretization effects kick in and the overlap and Wilson 
results start to differ. A comparison with the equivalent 
plot in \cite{biwi}, where the results from various other 
lattice Dirac operators were presented, shows that the 
discretization effects of the overlap operator at large $\mu$
are comparable to other formulations.  

\vfill
\section{Summary}

In this article we have analyzed the energy density of the 
overlap operator at finite chemical potential. Following \cite{blwe},
the sign function in the overlap was implemented through the spectral 
theorem using the analytic continuation of the sign into the complex 
plane. The subtracted energy density
$\epsilon(\mu) - \epsilon(0)$ was analyzed for finite and zero 
temperature lattices. Fits of the data show that the expected 
continuum behavior is approached. No trace of unphysical
$\mu^2/a^2$ terms was found. We conclude that 
overlap fermions with chemical potential \cite{blwe} provide both,
chiral symmetry and the correct description of fermions at
finite density.

\vskip7mm
\noindent
{\bf Acknowledgments:} 
We thank Leonard Fister, Gabriele Jaritz, Christian Lang, Stefan Olejnik,
Tilo Wettig, and Florian Wodlei for discussions and checking some of our 
calculations. This work is supported by the Slovak Science and Technology 
Assistance Agency under Contract No.\ APVT--51--005704, and the Austrian
Exchange Service \"OAD.

\end{document}